%% file: main.tex
\documentclass[graybox]{svmult}

\usepackage{mathptmx}        
\usepackage{helvet}          
\usepackage{courier}         

\usepackage{makeidx}         
\usepackage{graphicx}        
\usepackage{multicol}        
\usepackage[bottom]{footmisc}

\usepackage{url}             

\makeindex             

\begin{document}




\mainmatter
\include{karp}

\backmatter
\appendix


\end{document}

%% file: karp.tex
%
%
%

{ 

\title*{Robustness and uncertainty of direct numerical simulation under the influence of rounding and noise}
\titlerunning{{DNS under the influence of rounding and noise}}
\author{M. Karp \and N. Jansson \and S. Rezaeiravesh \and S. Markidis \and P. Schlatter}
\institute{M. Karp \and N. Jansson \and S. Markidis \at KTH Royal Institute of Technology, \email{{makarp,njansson,markidis}@kth.se} \and
S. Rezaeiravesh \at The University of Manchester, UK, \email{saleh.rezaeiravesh@manchester.ac.uk}
\and P. Schlatter \at Institute of Fluid Mechanics (LSTM), Friedrich--Alexander--Universit\"at (FAU) Erlangen--N\"urnberg, \email{philipp.schlatter@fau.de}}
\maketitle

\section{Introduction}
Numerical precision in large-scale scientific computations has become an emerging topic due to recent developments in computer hardware. Lower floating point precision offers the potential for significant performance improvements, but the uncertainty added from reducing the numerical precision is a major obstacle for it to reach prevalence in high-fidelity simulations of turbulence~\cite{witherden}. In the present work, the impact of reducing the numerical precision under different rounding schemes is investigated and compared to the presence of white noise in the simulation data to obtain statistical averages of different quantities in the flow. To investigate how this impacts the simulation, an experimental methodology to assess the impact of these sources of uncertainty is proposed, in which each realization $u^i$ at time $t_i$ is perturbed, either by constraining the flow to a coarser discretization of the phase space (corresponding to low precision formats) or by perturbing the flow with white noise with a uniform distribution. The purpose of this approach is to assess the limiting factors for precision, and how robust a direct numerical simulation (DNS) is to noise and numerical precision.

Evaluating numerical precision comes with a large set of challenges as the accumulation of round-off errors depends largely on the numerical method, grid points, and modeled parameters. As such, we constrain our focus to only evaluate the effect of the discretization of the state of the flow given some state vector $\mathbf{u}$ and the integration in time is performed through a chaotic map $\mathbf{F}(\mathbf{u})$. We consider a simulation to be represented by  the discretized system~$\mathbf{u}^{i+1} = \mathbf{F}(\mathbf{u}^i)$ where $\mathbf{u}^i =\{u^i_1,\ldots, u^i_n\}$ and $\mathbf{u} \in F^n$ where $F^n\in R^n$ is a finite precision discretization of the phase space $\mathcal{P}$. To assess the sensitivity and robustness of such a discretization we evaluate the impact of the following perturbed system~$
    \mathbf{u}^{i+1} = \mathbf{G}(\mathbf{F}(\mathbf{u}^i))$
where $\mathbf{G}(\mathbf{u}) = \{g(u_1),\ldots,g(u_n)\}$ maps an original state $\mathbf{u}$ to one in lower precision or impacted by noise. In this work, three different versions of the scalar functions~$g$ are considered:
$g(u) = \textrm{round}_{\rm{RTN,FP}}(u)$, $g(u) = \textrm{round}_{\rm{SR,FP}}(u)$, and  $g(u) = (1+\epsilon \delta)u$ where $\delta \sim U_{[-1,1]}$ (Noise case). Here, $\textrm{round}_{\rm{RTN,FP}}$ corresponds to conventional rounding-to-nearest (RTN) to a numerical representation in floating point precision (FP), $\textrm{round}_{\rm{SR,FP}}$ refers to stochastic rounding (SR) to precision FP, and in the latter uniformly distributed white noise is added  to~$u$ for the new value being sampled from $[1-\epsilon, 1+\epsilon]u$. 

Stochastic rounding for a value $x$ between two values $x_1$ and $x_2$ corresponds to rounding up to $x_2$ with probability $(x-x_1)/(x_2-x_1)$ and down to $x_1$ with probability $(x_2-x)/(x_2-x_1)$ while round-to-nearest would deterministically round $x$ to $x_1$ and $x_2$ whichever is closest to $x$. Incorporating stochastic rounding is motivated by recent works indicating that this method avoids the so-called stagnation in dynamic systems, where $x$ is repeatedly rounded to the same value, although $x$ changes~\cite{stocround}. 

For the floating point numbers considered in this work, we introduce some notation. A floating point number is uniquely defined by the number of mantissa bits~$b$, and exponent bits~$b_e$ together with one sign bit~$s$ dictate the sign of the floating point number. If we let~$e$ be the value of the exponent (as an unsigned~$b_e$-bit integer), and~$c_i$ for the $i$-th least significant bit of the mantissa, the value for a given floating point number is $
    (-1)^s \left(1 + \sum_{i=1}^{b} c_{b - i} 2^{-i}\right ) \times 2^{k - 1023}$,
where we have $b=52$ and $b_e=11$ in the case of IEEE double precision defined by 64 bits/8 bytes. In this work, we consider six different floating point formats ranging from IEEE double precision down to quarter precision. We list the properties of the considered number formats in Table \ref{tab:FP}. Especially important for our discussion will be the machine epsilon $\varepsilon$ which is the smallest relative difference a floating point number can represent.
\begin{table}[t]
\centering
    \begin{tabular}{lllll}
    \hline
        Name & Bits& $b$ & $b_e$ & {\centering $\varepsilon$}\\
        \hline
        FP64\rule{0pt}{2.4ex}& 64 & 52 & 11&$2^{-53}\approx 10^{-16}$  \\
        FP32 & 32 & 23 & 8&$2^{-24}\approx 6\cdot 10^{-8}$ \\
        FP16 & 16 & 10 & 5&$2^{-11}\approx 5\cdot 10^{-4}$ \\
        bfloat16 & 16 & 7 & 8 & $2^{-8}\approx 4\cdot 10^{-3}$  \\
        q43 & 8 & 3 &4 &$2^{-4} = 0.0625$  \\
        q52 & 8 & 2 & 5&$2^{-3} = 0.125$ \\ 
        \hline
    \end{tabular}
    \caption{Table of the different floating point formats considered ranging from FP64 to the recent FP8 formats q43 and q52.}
    \label{tab:FP}
\end{table}
A simulation is uniquely defined at time step $i$ by iterating the mapping $\mathbf{F}$ over $i$ time steps, letting $\mathbf{F}^i$ be the mapping applied $i$ times. The simulation can, as such, be uniquely represented by a series of states
$\mathbf{U}=\{\mathbf{u}^0,\mathbf{F}(\mathbf{u}^0),\ldots ,\mathbf{F}^i(\mathbf{u}^0)\} = \{\mathbf{u}^0,\mathbf{u}^1,\ldots,\mathbf{u}^i\},
$
or equivalently, by $\Delta \mathbf{U}= \{\Delta \mathbf{u}^1,\Delta \mathbf{u}^2,\ldots,\Delta \mathbf{u}^i\}$
with $\mathbf{u}^0$ being some initial condition and $\Delta \mathbf{u}^i = \mathbf{u}^i-\mathbf{u}^{i-1}$.
With this view, we consider two cases. First, the perturbed system $\hat{\mathbf{u}}^{i+1} = \mathbf{G}(\mathbf{F}(\hat{\mathbf{u}}^i)) = \hat\mathbf{F}(\hat \mathbf{u}^i)$ is obtained from perturbing the mapping repeatedly at each time step and the initial conditions, meaning the simulation can be defined as~$\hat{\mathbf{u}}^0=\mathbf{G}(\mathbf{u}^0)$ where $
\hat{\mathbf{U}}= \{\mathbf{G}({\mathbf{u}^0}),\hat{\mathbf{F}}(\hat\mathbf{u}^0)),\ldots ,\hat{\mathbf{F}}^i(\hat\mathbf{u}^0))\} = \{\hat{\mathbf{u}}^0,\hat{\mathbf{u}}^1,\ldots,\hat{\mathbf{u}}^i\}$. Second, the perturbed system is based on $\Delta \hat{\mathbf{u}}^i = \hat{\mathbf{u}}^i-\hat{\mathbf{u}}^{i-1}$ with $\Delta \hat{\mathbf{U}}$ defined similarly to $\Delta \mathbf{U}$.

We also study the impact of only rounding the observation of the state at each time step according to $\tilde{\mathbf{U}}=\{\mathbf{G}({\mathbf{u}^0}),\mathbf{G}(\mathbf{F}(\mathbf{u}^0)),\ldots ,\mathbf{G}(\mathbf{F}^i(\mathbf{u}^0)\} = \{\tilde{\mathbf{u}}^0,\tilde{\mathbf{u}}^1,\ldots,\tilde{\mathbf{u}}^i\}$
with $\Delta \tilde{\mathbf{U}}$ being based on $\Delta \tilde{\mathbf{u}}^i =\tilde{\mathbf{u}}^i-\tilde{\mathbf{u}}^{i-1}$. The idea of considering the system $\tilde{\mathbf{U}}$ is to provide a measure of how accurate the measurements themselves need to be without impacting the integration of the system in time.  In some sense, it is similar to how physical experiments are impacted by the precision of the sensors, with the difference that instead of physical measurement we evaluate $\mathbf{U}$ from~simulation. 

In this work, we only consider \textit{dynamical} systems that are not in a steady state but $\Delta \mathbf{U}$ is statistically stationary. In other words, the systems evolve in time meaning that the change in time $\Delta \mathbf{u}$ never completely stagnates. In addition, we focus on systems $\mathbf{F}$ that are ergodic and chaotic and the exact trajectory of the state is unknown. In other words, we assume a positive Lyapunov exponent of the system where an infinitesimal perturbation between two states causes the two trajectories to diverge exponentially, and conventional error analysis is difficult to perform. The goal in our case is to carry out a simulation for a long enough time to cover the phase space and explore an attractor of interest. By computing statistics, after a statistically stationary state has been reached, we want to determine the dynamics and probability density function (PDF) of the attractor in a turbulent regime and see how they are affected by numerical precision.

\section{Lower precision, rounding, and noise in turbulent channel flow}
We carry out a number of simulations of turbulent channel flow at $Re_\tau=180$ (friction-based Reynolds number) similar to our previous work in~\cite{karp23}, but rather than focusing on reducing the number of mantissa bits we utilize different established IEEE floating point formats. An important note is that the dimensionality and number of degrees of freedom of the state $\mathbf{u}$ and the correlation dimension in a simulation of turbulence is very large. In particular, the correlation dimension~$d$ scales with the Reynolds number, and as the probability of arriving in a periodic orbit at a given time step scales as $\varepsilon^{-d/2}$ for a finite precision $\varepsilon$, this probability quickly becomes very small~\cite{correlation}. However, the state is instead prone to stagnation where $\Delta \mathbf{u}$ becomes zero and the numerical precision fails to resolve the change from one time step to another, this further amplifies the importance of looking at the PDFs of the state change in time, $\Delta \mathbf{U}$, in addition to the state itself.

For our analysis, we use simulations performed with the spectral element framework Neko~\cite{neko}. 
All quantities are non-dimensionalized by the half-channel height~$\delta$ and the bulk velocity~$U_b$. We consider the baseline simulation $\mathbf{U}$ to be the one carried out in full FP64. $\hat\mathbf{U}$ are the simulations where the state is either perturbed at each time step to a lower floating point precision (rounded either with SR or RTN) or contaminated by adding noise. $\tilde{\mathbf{U}}$ is when only the observation of the state is perturbed. We consider the precisions listed in Table~\ref{tab:FP} and emulate the precisions with the CPFloat library~\cite{cpfloat}. In our plots, we show $\hat{\mathbf{U}}$ and $ \tilde{\mathbf{U}}$, omitting $\mathbf{U}$ as corresponding results were not different from those of $\hat{\mathbf{U}}$ or $\tilde{\mathbf{U}}$ in FP32.
\begin{figure}[t]
    \centering
    \includegraphics[width=0.47\columnwidth]{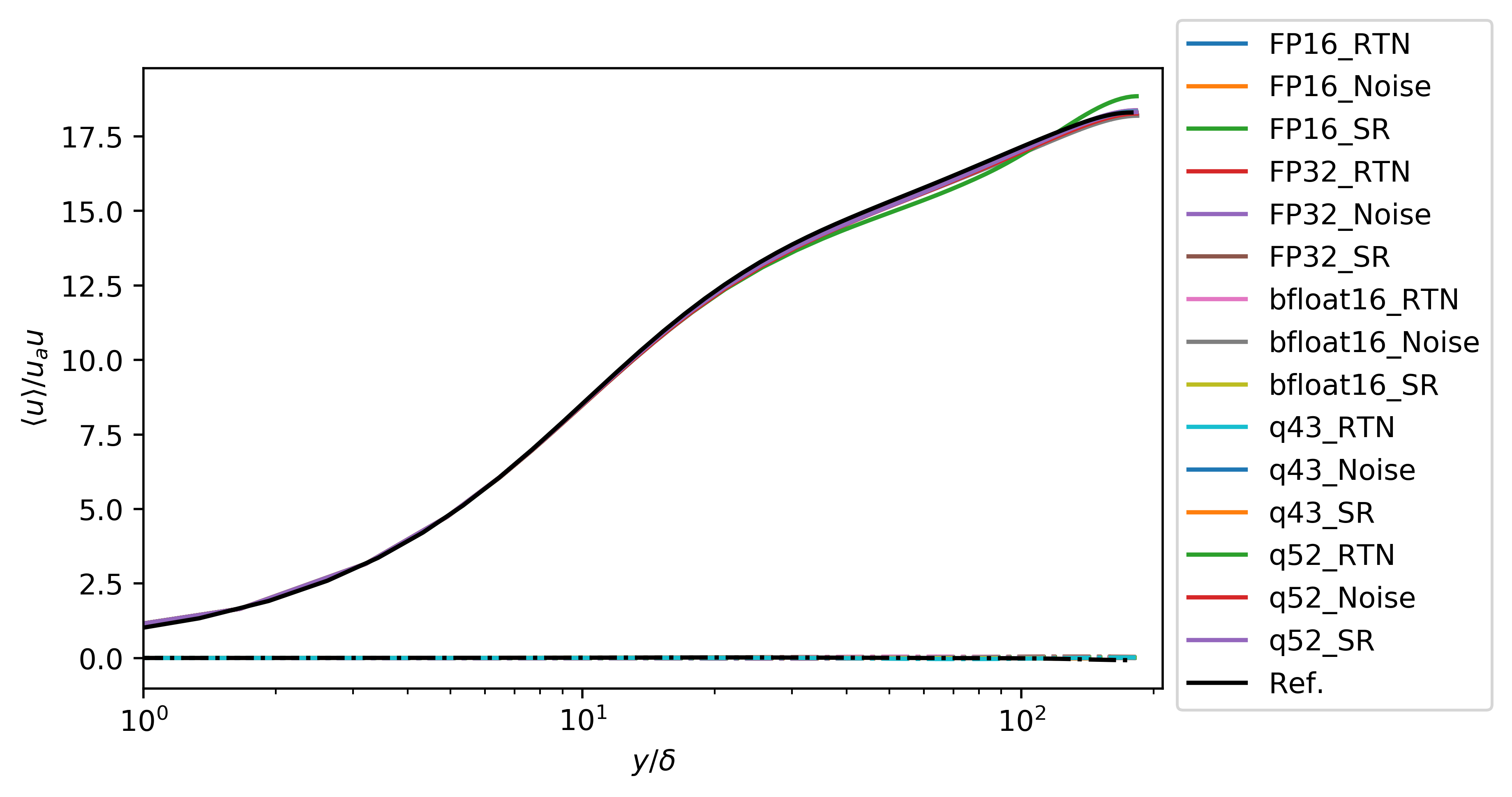}
    \includegraphics[width=0.47\columnwidth]{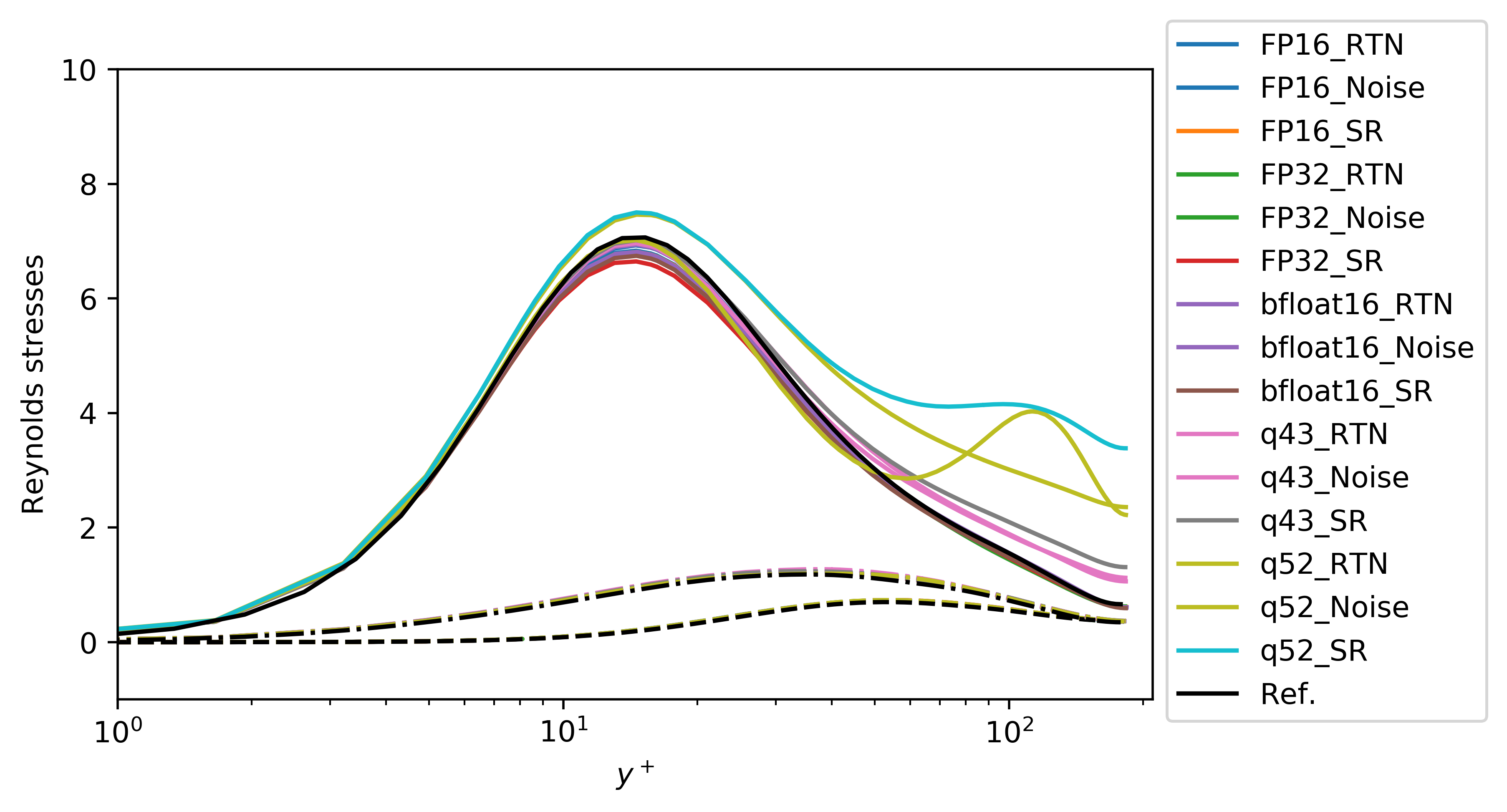}
    \includegraphics[width=0.47\columnwidth]{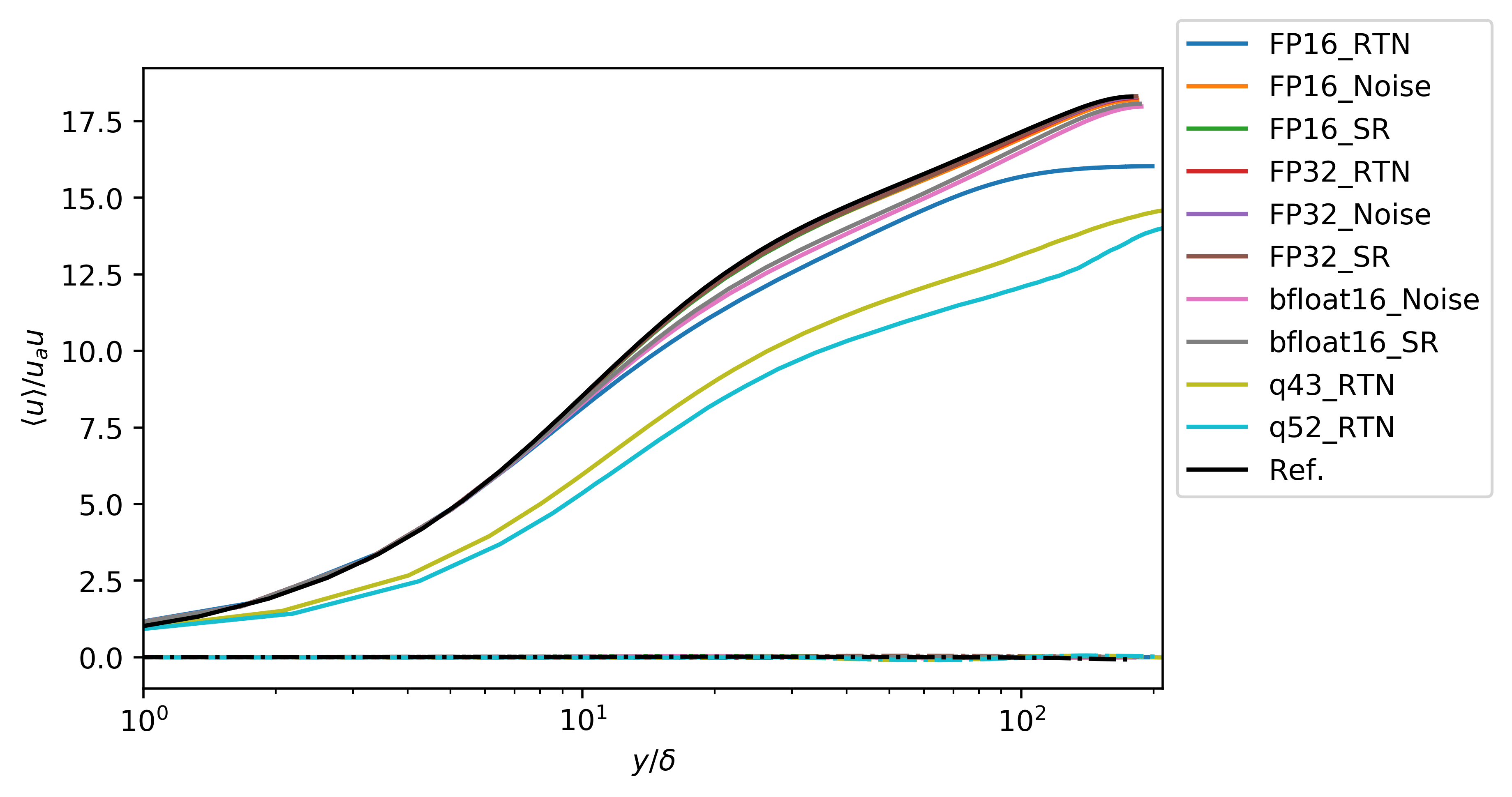}
    \includegraphics[width=0.47\columnwidth]{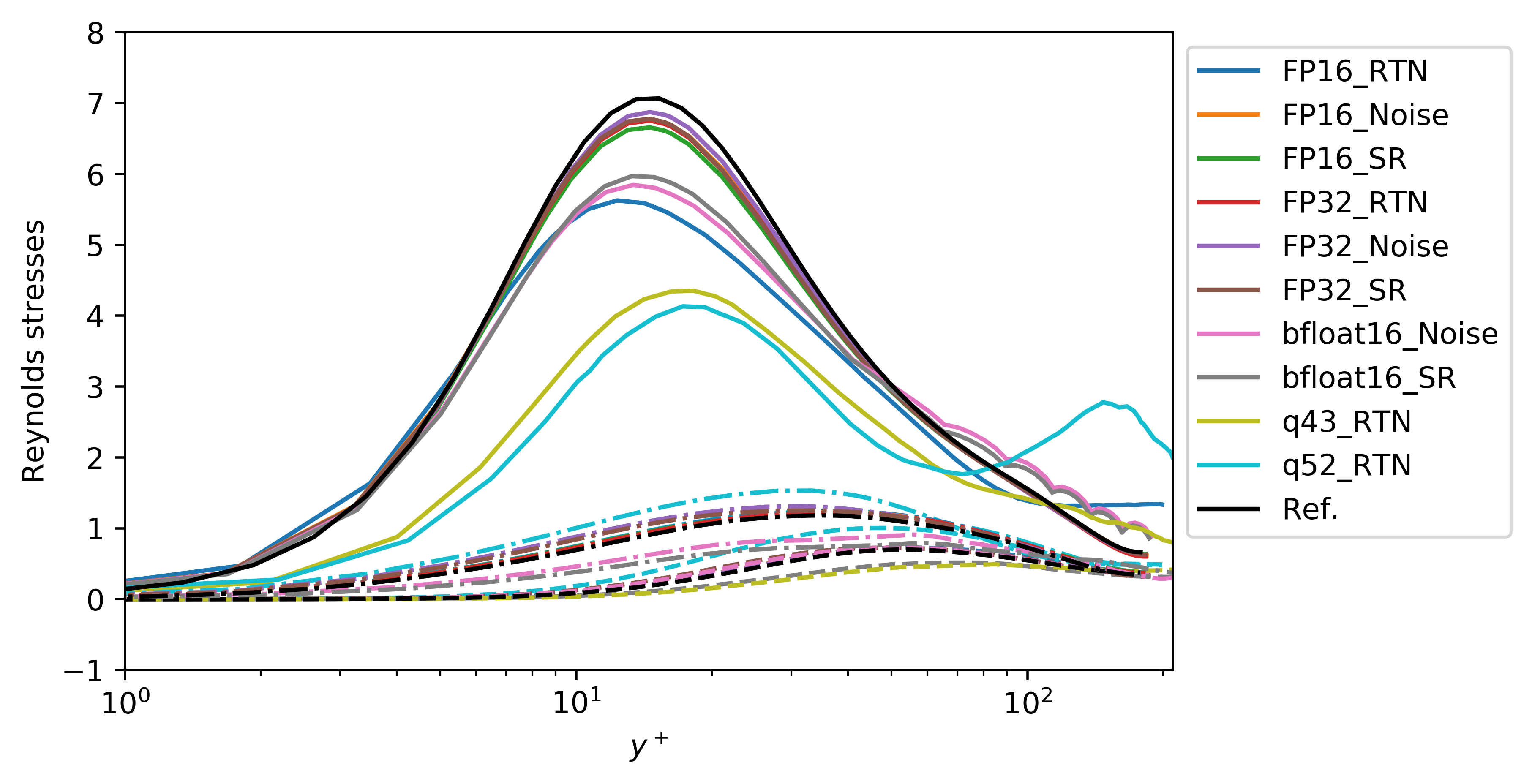}
    \caption{First- (left) and second-order (right) statistical moments of velocity $\tilde{\mathbf{U}}$ (top) and $\hat{\mathbf{U}}$ (bottom). 
    The solid, dashed and dot-dashed lines correspond to the streamwise~($u$), wall-normal~($v$), and spanwise~($w$) velocity components, respectively. Note that only isotropic components of the Reynolds stress tensor are plotted (right).  
    }
    \label{fig:stats}
\end{figure}
The profiles of the first and second-order statistical moments of $\tilde{\mathbf{U}}$ and $\hat\mathbf{U}$ are shown in Figure \ref{fig:stats} and compared to reference data from~\cite{moser}. The cases with bfloat16\_RTN and the quarter precision formats with SR and Noise did not complete for the $\hat\mathbf{U}$ cases.
A general observation is that the RTN method performs worse compared to the SR and Noise methods. 
The mean profiles of $\tilde{\mathbf{U}}$, where only the observation is perturbed, are only affected by very low precisions, such as q52\_RTN, where considerable bias is introduced for each sample.
Especially, for the SR and Noise cases, the sample-estimated mean is still maintained, and only the time required for the statistics to converge would be impacted. However, for the second-order moments, the computed statistics are more sensitive to perturbations, as the artificial fluctuations induced impact the results, even when using SR and Noise methods. For q43, these indicate that even sampling in quarter precision is insufficient. In contrast, sampling in half-precision is able to capture the correct trend.

For the perturbed system $\hat{\mathbf{U}}$, the observations are slightly different. In particular, for the lowest precisions, the solution either stagnates for too low precision (RTN), or the sharp gradients appearing in the solution would crash the simulation (SR and Noise). While the impact of noise and SR for bfloat16 on the mean flow is not as significant compared to the RTN, the Reynolds stresses show more sensitivity. Of note, is that FP16 with SR or Noise once again performs similarly to the simulations with higher precisions. Our results for FP16\_RTN are very similar to the observed behavior as in our previous work~\cite{karp23} where~$\hat{\mathbf{U}}$ was rounded with RTN to fewer mantissa bits while keeping the exponent constant. In that study, it was found that the point at which the time series differed was around $b=10$, the same number of bits as FP16. That FP16\_RTN also impacts the computed statistics is clearly reflected in $\hat{\mathbf{U}}$ for the Reynolds stresses. However, the computed statistics do not yield information about why exactly the limit is reached around $b=10$ for RTN. For this, we consider the PDFs of $\Delta \mathbf{U}$ scaled by the mean flow velocity, and $\mathbf{U}$. 

\begin{figure}[t]
    \centering    \includegraphics[width=0.49\columnwidth]{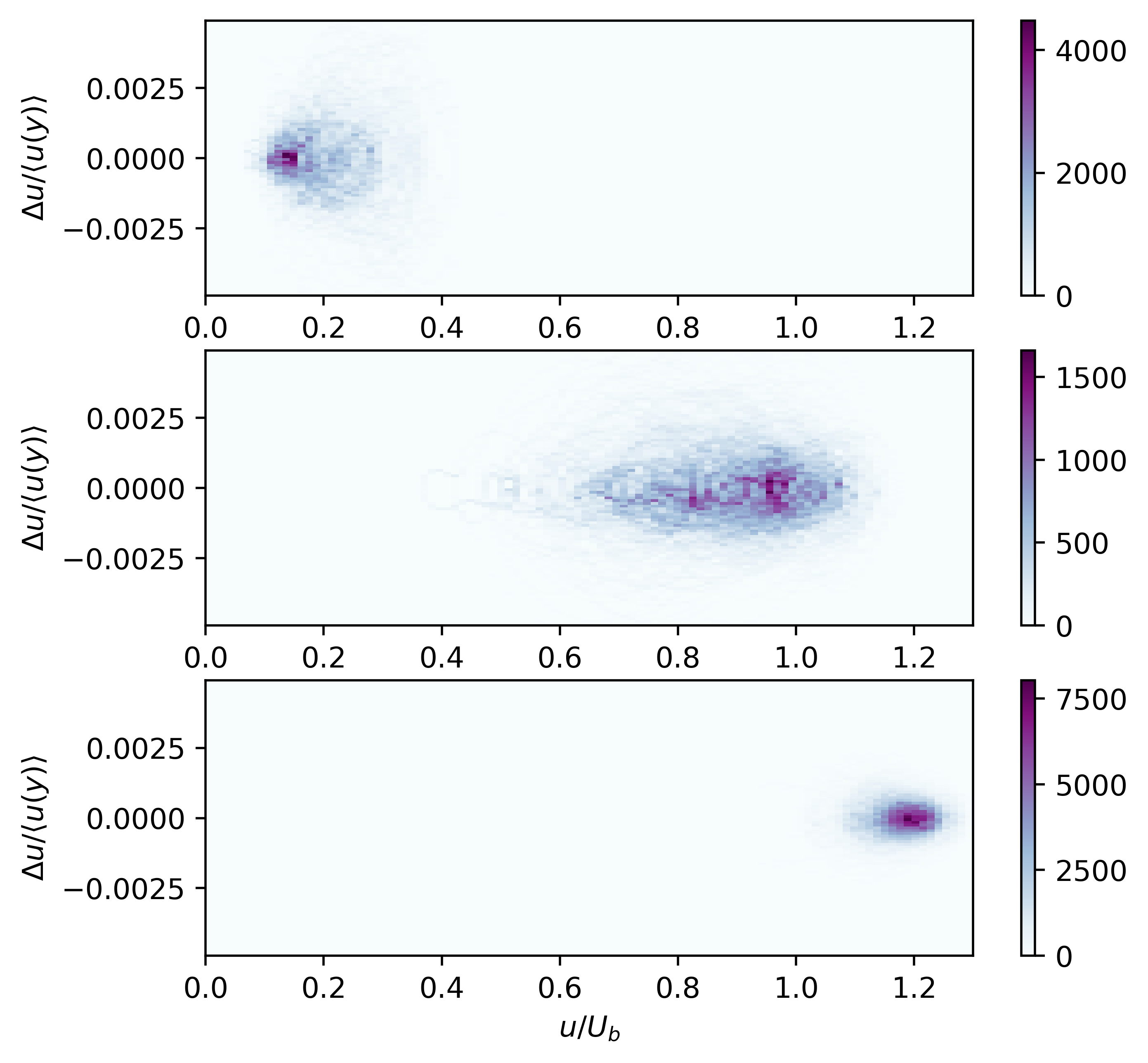}
    \includegraphics[width=0.49\columnwidth]{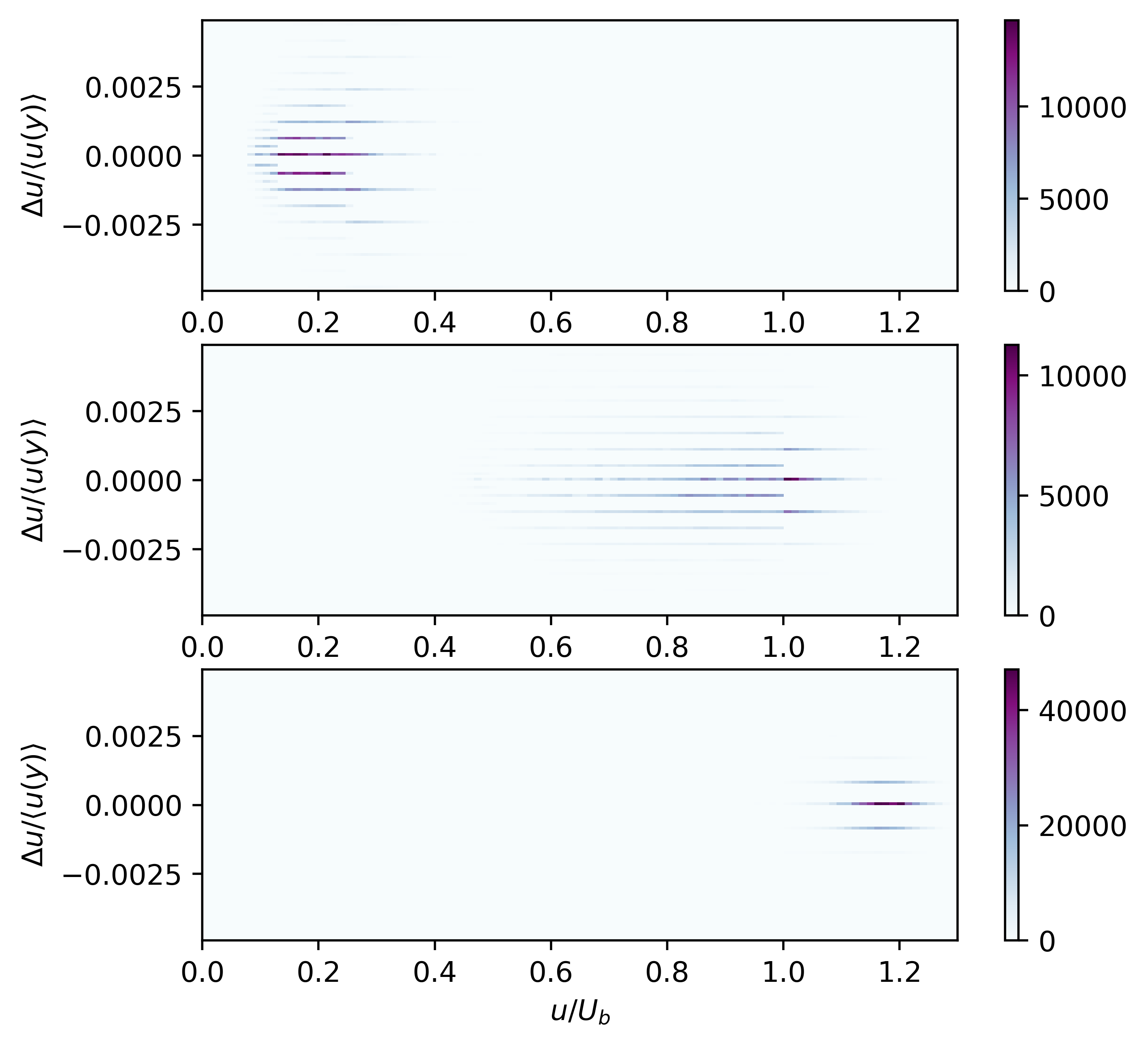}
    \includegraphics[width=0.49\columnwidth]{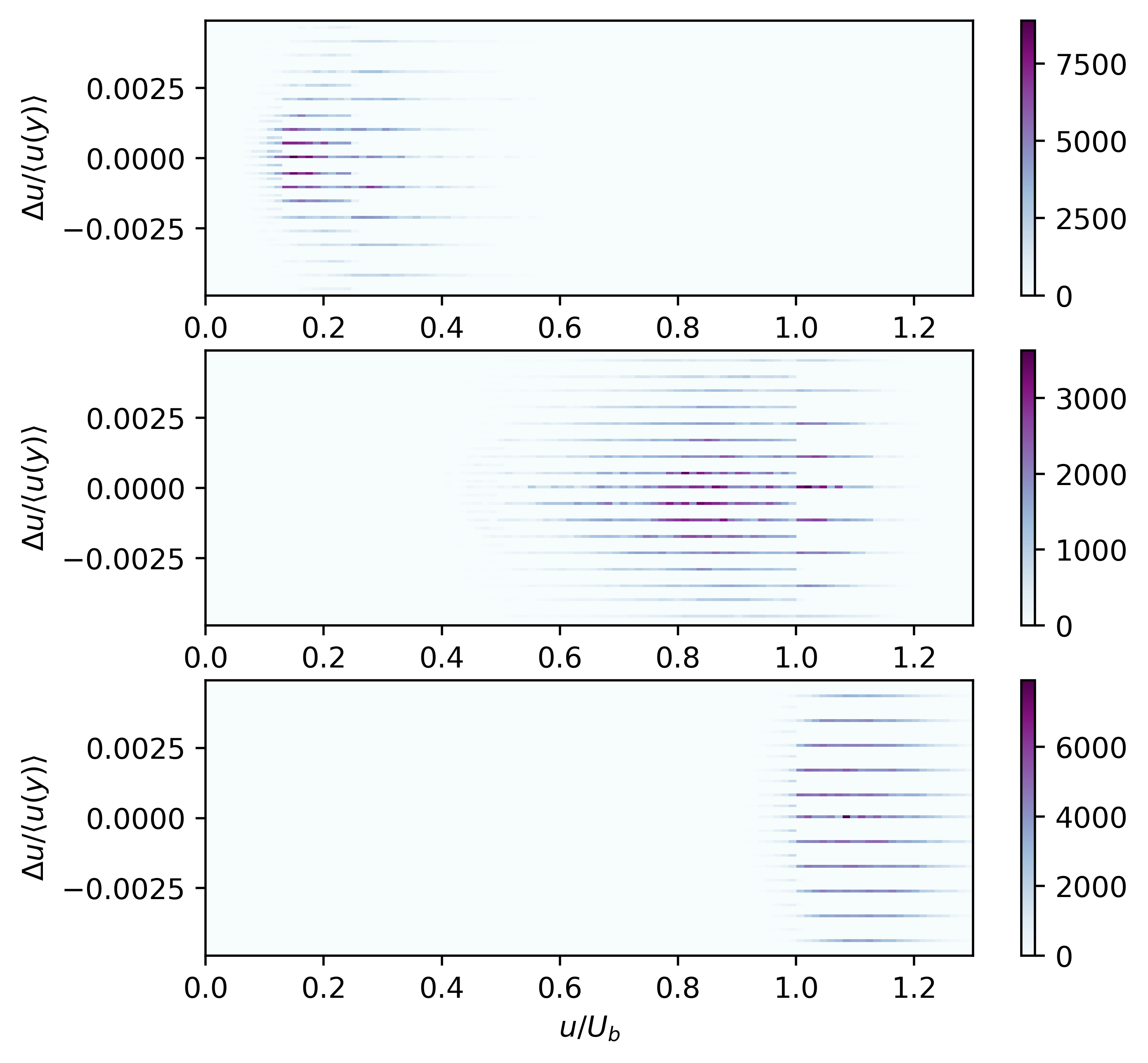}
    \includegraphics[width=0.49\columnwidth]{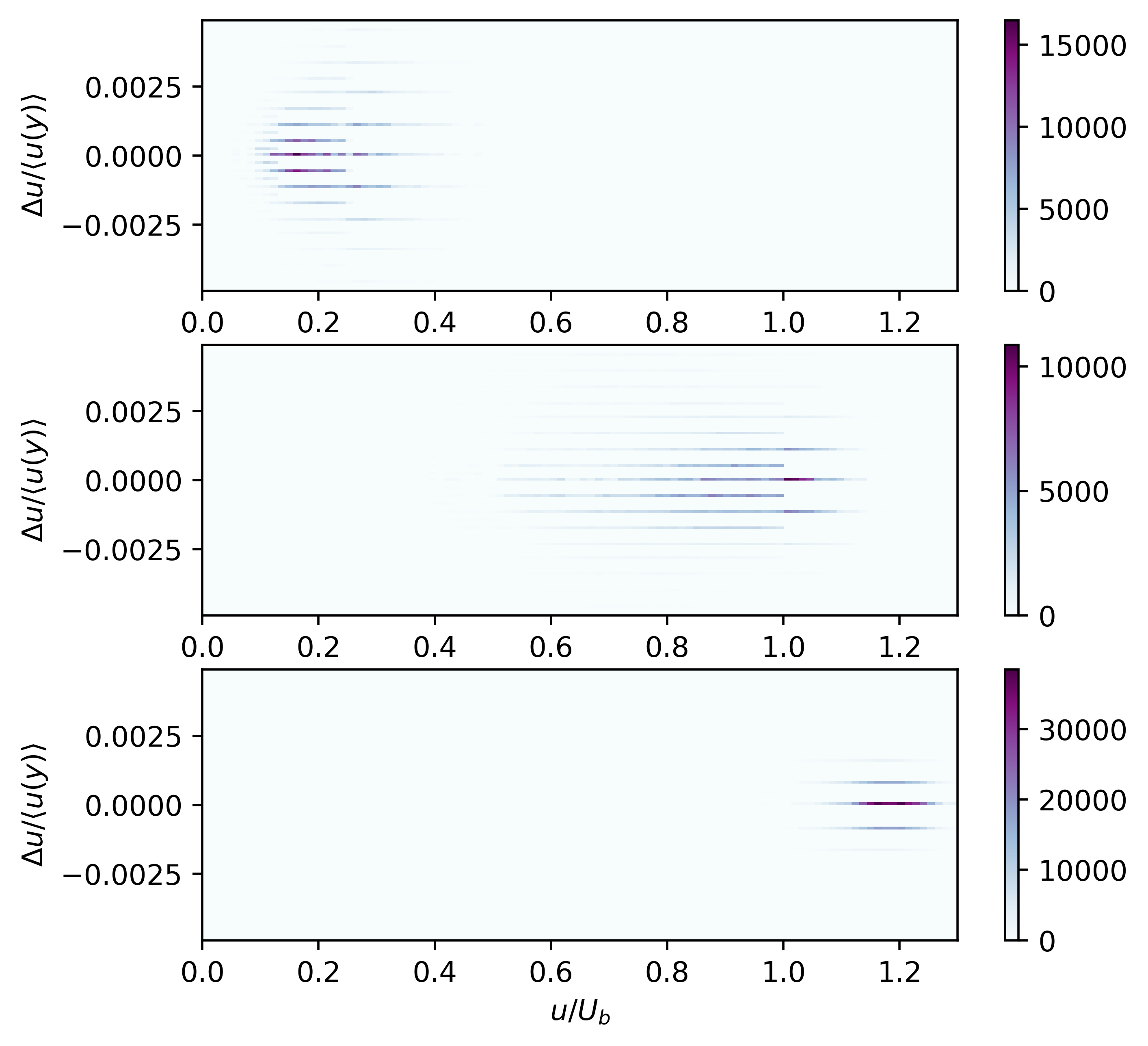}
    \caption{2D PDFs of the streamwise velocity component $u$ and $\Delta u = u^i -u^{i-1}$ normalized by the mean velocity at three different wall-normal locations $y=\{0.0175, 0.14, 0.92\}$. The three different locations are shown for $\tilde{\mathbf{U}}$ (top) in FP32 (left) and FP16 (right) together with the results from $\hat{\mathbf{U}}$ (bottom) with FP16 RTN (left) and FP16 SR (right)}
    \label{fig:2dpdfs_simcomp}
\end{figure}
As we perform the rounding in time, it is natural to evaluate how the state changes for different variables in $\mathbf{u}$. We show the joint PDFs of $\mathbf{u}$ and $\Delta \mathbf{u}$  in Figure~\ref{fig:2dpdfs_simcomp} for FP32 and FP16 for both rounding the simulation and only the sampling of statistics. If we first consider FP32, it is clear that the distribution of~$u$ is significantly more skewed than that of the velocity change, which exhibits a more symmetric behavior. Another important note is the spread of the distributions, where~$u$ varies a lot more in the log layer compared to the mean velocity at the same point compared to the near-wall region and the center of the channel. As for the characterization of $\Delta u$ an important aspect is that the relative velocity change and turbulence intensity is a lot smaller closer to the center. When we sample the field in lower numerical precision, the distributions of the locations closer to the wall are captured by significantly more values for FP16 than in the center of the channel. This is coherent with our findings in~\cite{karp23} where we noted that the near-wall region was less impacted by changing the state to a lower floating point precision. 

Evaluating $\hat{\mathbf{U}}$ and $\Delta \hat{\mathbf{U}}$ instead for FP16 with RTN and SR, we obtain the joint PDFs shown at the bottom of the figure. In this plot, the steep gradients introduced by using deterministic rounding are clearly visible. Especially for the distribution close to the center line, the distribution is significantly altered. Using SR alleviates this problem and the similarity between only sampling in FP16 and computing the actual solution with SR is very similar. The main differences in the plots are due to the slightly different lengths of the time series that leads to a different scaling of the colormap.
However, this is only when we look at the relative velocity change. Considering that the smallest scales in the flow are constant, they will be relatively smaller compared to the mean velocity in the middle of the channel. This indicates that conventional floating point formats might not be the best way to represent the state vector in turbulent flows when using fewer bits. Instead, perhaps a fixed precision format with spacing based on the standard deviation in the near-wall region would yield better results, as such a format would properly capture the fluctuations and changes in the middle of the channel as well. While floating point precision is a format with enormous range and flexibility it is an open question if another numerical format would be able to better capture the dynamics between the different scales in turbulent flows. Our results indicate that for DNS, where we want to resolve the PDF of the velocity change in time, the floating point precision is especially important in regions with small relative velocity changes and low turbulence intensity. Going forward, the interaction between the flow physics and simulation parameters and also how they impact the required precision warrants further investigation.

}